\documentclass[%
 reprint,
amsmath,amssymb,aps,
]{revtex4-1}

\usepackage{lipsum}
\usepackage{graphicx}
\usepackage{dcolumn}
\usepackage{bm}

\usepackage{color}
\usepackage{xspace}

\usepackage{hyperref}

\def\pdef{$+\frac{1}{2}$\xspace}
\def\mdef{$-\frac{1}{2}$\xspace}
\def\pmdef{$\pm\frac{1}{2}$\xspace}

\newcommand{\fig}{Fig. }
\DeclareMathOperator{\Tr}{Tr}

\makeatletter
\newcommand{\printfnsymbol}[1]{%
  \textsuperscript{\@fnsymbol{#1}}%
}
\makeatother

\begin{document}

\title{Self-organized dynamics and the transition to turbulence of confined active nematics}
\author{Achini Opathalage}
\email{AO and MMN contributed equally to this work.}
\author{Michael M. Norton}
\email{AO and MMN contributed equally to this work.}
\author{Michael P. N. Juniper}
\author{S. Ali Aghvami}
\author{Blake Langeslay}
\author{Seth Fraden}
\email{fraden@brandeis.edu}
\affiliation{Department of Physics, Brandeis University, Waltham, MA 02453}

\author{Zvonimir Dogic}
\email{zdogic@physics.ucsb.edu}
\affiliation{Department of Physics, University of California at Santa Barbara, Santa Barbara, CA 93106}


\date{\today}
\begin{abstract}
We study how confinement transforms the chaotic dynamics of bulk microtubule-based active nematics into regular spatiotemporal patterns. For weak confinements, multiple continuously nucleating and annihilating topological defects self-organize into persistent circular flows of either handedness. Increasing confinement strength leads to the emergence of distinct dynamics, in which the slow periodic nucleation of topological defects at the boundary is superimposed onto a fast procession of a pair of defects. A defect pair migrates towards the confinement core over multiple rotation cycles, while the associated nematic director field evolves from a distinct double spiral towards a nearly circularly symmetric configuration. The collapse of the defect orbits is punctuated by another boundary-localized nucleation event, that sets up long-term doubly-periodic dynamics. Comparing experimental data to a theoretical model of an active nematic, reveals that theory captures the fast procession of a pair of $+\frac{1}{2}$ defects, but not the slow spiral transformation nor the periodic nucleation of defect pairs. Theory also fails to predict the emergence of circular flows in the weak confinement regime. The developed confinement methods are generalized to more complex geometries, providing a robust microfluidic platform for rationally engineering two-dimensional autonomous flows. 
\end{abstract}

\maketitle



\section*{Introduction}

Powered by the consumption of chemical energy, active nematic liquid crystals generate mesoscopic active stresses that render the entire system unstable. For extensile nematics slight bend distortions of the director field create hydrodynamic flows that further deform the director field~\cite{Simha2002,Marchetti2013, ramaswamy2010mechanics, shelley2016dynamics}. Upon saturating, these distortions produce pairs of motile topological defects that drive large-scale turbulent-like dynamics~\cite{Giomi2013, Thampi2014c,Gao2015d,Giomi2015,shankar2018defect,Thampi2013, Gao2015d,Oza2016,narayan2007long,Zhou1265,Sanchez2012,Decamp2015,ellis2018curvature,guillamat2016control}. Design of active matter based devices that directly convert chemical energy into macroscopic mechanical work requires predictive control of the emergent spatiotemporal patterns. In passive liquid crystals such goals can be accomplished by prescribing confinement geometry and boundary anchoring conditions, which in turn determine the configuration of the director field throughout the sample~\cite{Luo2016,Senyuk2013,Peng2016}. However, the full extent to which the dynamics of active liquid crystals can be prescribed through control of the boundaries remains an open question. For example, recent theoretical work suggests that circular confinement generate robust circular flows that are largely independent of the details of the director boundary conditions~\cite{Norton2018a}. 

\begin{figure*}
\includegraphics[width=\textwidth]{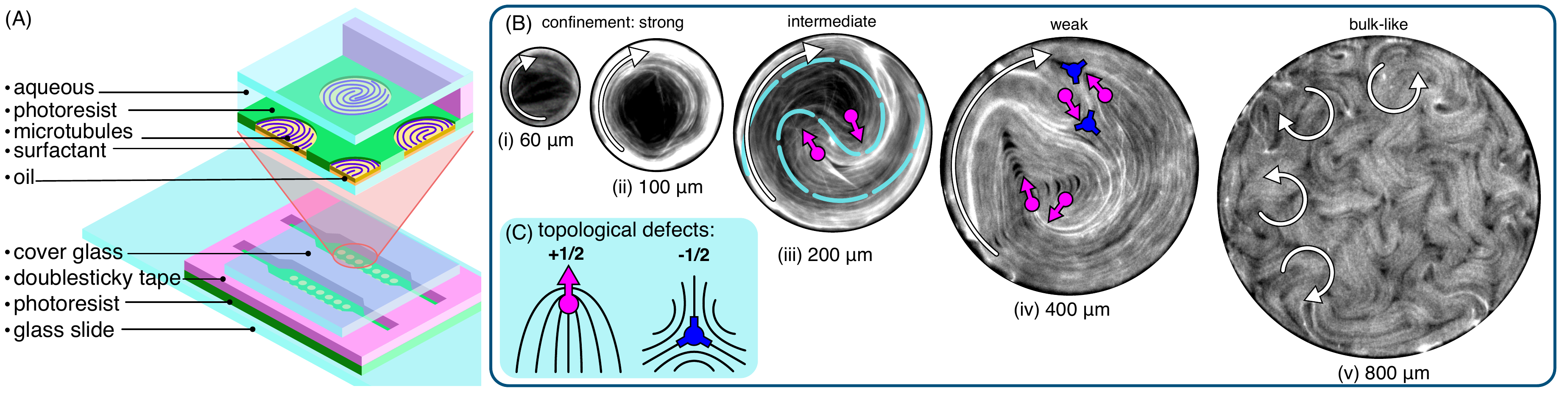}
\caption{\textbf {Active nematics in weak, intermediate and strong confinement regimes. (A)} Schematic of sample cell used to confine active nematic. A micro-pattern of cylindrical holes was imprinted into a photoresist layer bound to a microscope slide. Oil filled flow chamber was displaced with an aqueous active mixture, leaving behind holes with a circular oil-water interface located near the top of the cylindrical opening. Centrifuging microtubules onto the oil-water interface led to the formation of confined active nematic. \textbf{(B)} (i-v) Fluorescence images of confined active nematics with diameter ranging from 60 to 800 $\mu$m. White arrows indicate direction of circulation. Cyan line overlaying 200$\mu$m disk highlights the double spiral configuration of the nematic director observed for intermediate confinements. Photographs are not to scale. \textbf{(C)} Structure of nematic director field around topological defects of charge \pdef (magenta) and \mdef (blue).}
\label{figure1}
\end{figure*}

The two-dimensional active liquid crystals studied here are formed by depositing microtubule fibers comprised of individual microtubules microns in length bundled together by depletion forces on an oil-water interface. Clusters of ATP-powered kinesin motors bind to neighboring microtubules powering their sliding and generating net extensile stresses~\cite{Sanchez2012,Decamp2015}. We confined 2D active nematics by filling photolithographically shaped holes formed in photoresist with oil and covering the oil interface with an aqueous suspension of active fibers. The nematic thus formed is a thin 2D active material with in-plane boundaries provided by the processed photoresist (\fig\ref{figure1}). The hard-wall boundaries enforce both the no-slip condition and the parallel anchoring that are readily described by hydrodynamic models of confined active nematics~\cite{Woodhouse2012, Shendruk2017,Gao2017,Norton2018a,theillard2017geometric,chen2018dynamics}. 

Below a critical confinement size we observe self-organized circular flows, which are a ubiqutous feature of diverse experimental systems~\cite{Wioland2013, Lushi2014, Segerer2015, Wioland2016,Wu2017}. However, experimentally identified criteria for the onset of the circular flows are different from theoretical predictions. For a subset of confinements we observe slow and highly periodic dynamics of boundary induced defect nucleation that is superimposed on the fast circular flows; a phenomena we refer to as doubly-periodic dynamics. Related dynamics were also observed by imposing a distinct boundary condition in which the substrate has anisotropic viscosity due to a smectic substrate~\cite{guillamat2017taming}. Taken together these observations suggest that doubly-periodic dynamics are a ubiquitous feature of confined active nematics that is not accounted for by the current theoretical models. In continuum theories of passive and active low molecular weight liquid crystals, defects can propagate through molecular reorientation, a mechanism which does not move material. In contrast, here, defects avoid crossing material lines defined by the director field because their motion is heavily constrained by the long spatial extent of the constituent fibers. Our results on confined active nematics challenge existing theoretical models with experimental data, identifying physical features whose inclusion into theoretical models may reconcile the observed differences.  

\section*{Experimental Results}
Using microfluidic technologies we confined active nematics into circular geometries whose diameter ranged from 60 to 800 $\mu$m (\fig\ref{figure1}). For all confinements studied, microtubules exhibited parallel alignment to the boundaries, which ensured that the nematic director field contained topological defects. Specifically, circular geometry has an Euler characteristic of +1, thus the topological charge of all disclinations has to add up to this value~\cite{kamien2002geometry}. For the strongest confinements ranging up to 100 $\mu$m, the disk diameter was less than the defect size in a bulk nematic (\fig\ref{figure1}Bi,ii). Defect size indicates the minimum radius of curvature microtubule fibers can support without fragmenting; therefore in this regime, microtubules accumulate at the boundary leading to significant density gradients. For intermediate confinements (100 to 200 $\mu$m), the active nematics formed two \pdef defects typically arranged into an asymmetric double spiral configuration (\fig\ref{figure1}Biii). This is distinct from passive nematics in which the director field features two static \pdef defects oriented along the same line in an antipodal symmetric configuration~\cite{garlea2016finite,Duclos2016}. For weak confinements (300 to 600 $\mu$m) we observed proliferation of additional defects throughout the active nematic (Fig.\ref{figure1}Biv). In all three regimes active nematics exhibited robust circular flows. Further increasing the confinement size to 800 $\mu$m and beyond, yielded active nematics with turbulent bulk-like dynamics (\fig\ref{figure1}Bv). The constraint that the total topological charge adds up to +1 remained intact for all confinement regimes, since nucleation events only create topologically neutral \pmdef defect pairs.  

\begin{figure*}[!ht]
\includegraphics[width=\textwidth]{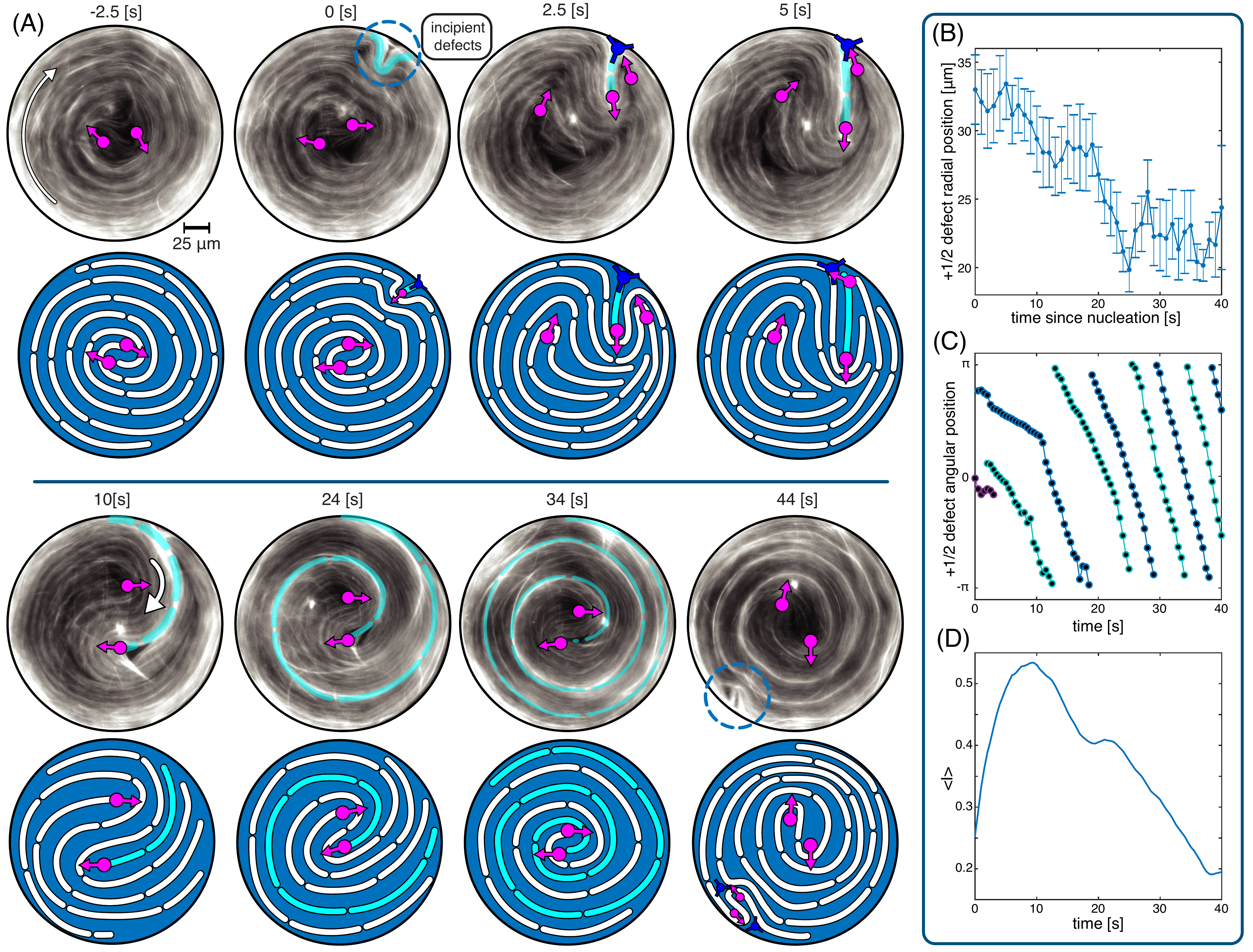}
\caption{\textbf{Dynamics of confined active nematic at intermediate confinements} \textbf{(A)} Fluorescence images and accompanying illustrations depict the spatiotemporal evolution of the nematic field during one cycle consisting of boundary induced defect nucleation and defect orbit collapse. (-2.5 s) The system begins with microtubules wound tightly into a spiral such that they appear to be arranged in closed, concentric circles. (0 s) Microtubules adjacent to the boundary (cyan) buckle, creating an inwardly propagating \pdef defect. (2.5 s - 5 s) A fracture forms between the \mdef defect created during nucleation and one of the original \pdef defects, returning the system to a  two-defect configuration. (10 s - 34 s) A pair of orbiting \pdef defects approach each other over multiple rotations. (44 s) The confined active nematic returns to the original, tightly wound and therefore almost perfectly circular configuration with weak azimuthal gradients. Over the next few seconds the microtubule layer once again buckles. A subsequent cycle is shown in \fig S2 and dynamics are shown in movie S2. {\bf (B)} Radial position of \pdef defects as a function of time shows the inward defect migration. The defect position was averaged over twenty cycles. The radial position increases towards the end of the cycle as buckling begins to push the existing defects away from the center. Error bars are standard error. {\bf (C)} Angular position of two \pdef defects during one defect nucleation cycle is indicative of the fast procession with a $\sim$ 10 s period, colors indicate different defects.  {\bf (D)} Fluorescence intensity, averaged over the central area with 50$\mu$m diameter. Defect nucleation repopulates the disk interior with microtubules for first 10 s; afterwards the microtubule density steadily decreases as the the defects circulate and filaments are expelled towards the boundary (10 s - 40 s). Confinement diameter is 200 $\mu$m.}
\label{figure2}
\end{figure*}

\begin{figure*}
\begin{center}
\includegraphics[width=\textwidth]{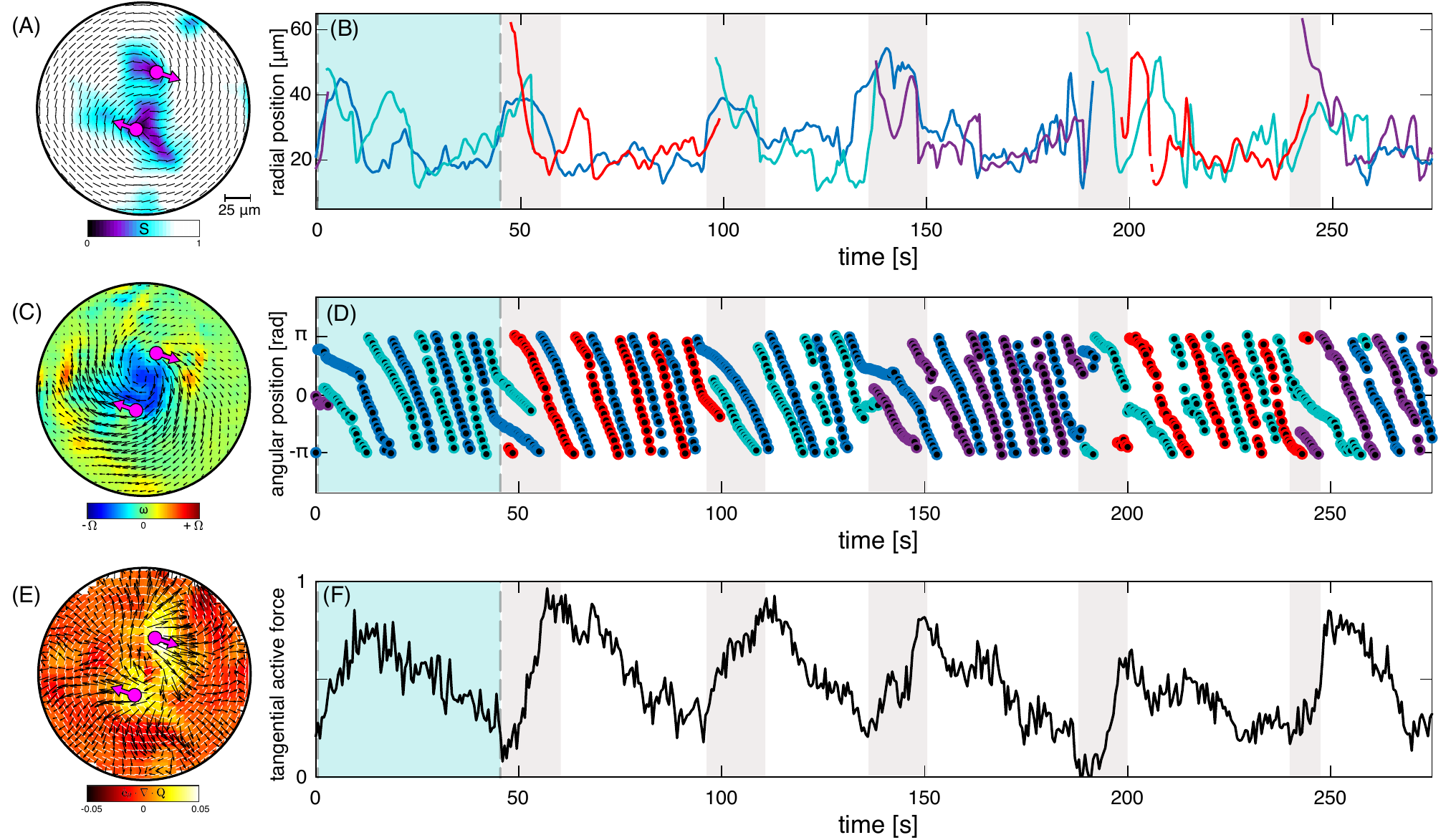}
\end{center}
\caption{\textbf{Long-time doubly-periodic dynamics at intermediate confinements.} {\bf (A)} Map of the director field $\mathbf{n}$ and the associated order parameter $S$ extracted from the fluorescence images. Two \pdef defects are also indicated with magenta arrows. {\bf (B)}  Radial position of \pdef defects plotted over several boundary nucleation events (period $\sim$ 50 sec) and many defect orbit periods (period $\sim$ 10 sec). Colors indicate different defects. {\bf (C)} PIV analysis yields the autonomous flow and the vorticity field (vorticity scale $\Omega$ = $\pm$ 0.8 sec$^{-1}$ ). {\bf (D)} Angular position of \pdef defects. The unchanging sign of the slope indicates that handedness of the circulation persists over multiple cycles of defect nucleation. Colors indicate different defects. {\bf (E)} Spatial map of $\nabla\cdot\mathbf{Q}$ (black arrows) and the magnitude of tangential component $\hat{\mathbf{e}}_{\theta}\cdot\nabla\cdot\mathbf{Q}$ (color field). {\bf (F)} Spatially integrated total active force in the tangential direction $\int {\textrm{d}\mathbf{x}\left(\hat{\mathbf{e}}_{\theta}\cdot\nabla\cdot\mathbf{Q}\right)}$ as a function of time. Maxima correspond to configurations where the two \pdef defects are far from one another, minima occur directly before nucleation when defects have effectively merged. Scale is normalized by 64 $\mu$m. Images in panels {\bf(A)}, {\bf(C)}, and {\bf(E)} correspond to \fig\ref{figure2}A (t=10 s). The light blue shaded window in {\bf(B)},{\bf(D)},{\bf(F)} corresponds to snapshots in \fig\ref{figure2}A. The grey shaded windows indicate nucleation events. Confinement diameter is 200 $\mu$m.}
\label{figure3}
\end{figure*}

\subsection*{Intermediate confinements}  We first describe dynamics in the intermediate confinement regime, which exhibited persistent circulating flows accompanied by complex but highly periodic dynamics of topological defects (Movie S1, S2). From a time sequence of fluorescently acquired images we extracted the spatiotemporal evolution of the nematic director field (\fig\ref{figure2}A), and using a previously developed algorithm we identified the locations of the topological defects~\cite{Decamp2015}. Immediately before a boundary nucleation event, the two circulating \pdef defects almost merged at the center, and the director field adopted a nearly circularly symmetric configuration (-2.5 s in Fig.\ref{figure2}A). This configuration was unstable and the effective defect merger at the confinement core coincided with initiation of a remote nucleation event at the boundary (0-5 s in \fig\ref{figure2}A). Specifically, over the next few seconds, the boundary-adjacent microtubule layer underwent a large scale inward buckling deformation that was driven by the intrinsic bend instability of the extensile active nematics. As the instability grew it generated a pair of \pmdef defects. The \pdef defect was propelled towards the center, while the oppositely charged \mdef defect remained at the boundary, eventually merging with one of the two original \pdef defects  (5-10 s in \fig\ref{figure2}A). The boundary nucleation event briefly disrupted the regular procession of the defects, but within a short time the active nematic reestablished the spiral configuration that powered robust circular flows. The periodicity of these flows was $\sim10$ s, as can be seen by plotting the angular position of the \pdef topological defects (\fig\ref{figure2}C). Over multiple rotations the defect pair arranged in a double spiral, slowly migrated towards the center (\fig\ref{figure2}B). This inward winding reconfigured the nematic director field, recreating the nearly circularly symmetric configuration (44 s in \fig\ref{figure2}A), which initiated another boundary induced nucleation event.  

The above described periodic dynamics were coupled to spatiotemporal changes in the local microtubule density. In accordance with theory that predicts the motion of extensile active nematics from regions of high to low curvature~\cite{Simha2002, Putzig2014}, the elongating microtubule fibers drove material toward the boundary leaving the interior devoid of nematic material (\fig\ref{figure2}D). In the subsequent cycle the disk's center was repopulated with filaments as the \pdef defect detached from the boundary and moved towards the center. This defect pulled along a bright fiber of microtubules, which served as a tracer for the deformations undergone by the nematic material elements (cyan line in \fig\ref{figure2}A). This fiber continuously elongated, winding around the center multiple times; from 0-44 s  the material element grew from a 10 $\mu$m protrusion to a several millimeter long coiled fiber. Notably, the two defects arranged at the core of the double spiral never crossed the elongating fiber and were directed inward in slowly collapsing orbits. 

Following the system on longer time scales revealed that the above described dynamics repeated itself with a period of $\sim 40 \pm 5$ s (\fig\ref{figure3}A, B). Time evolution of the angular defect position confirmed that the slow dynamics of boundary induced nucleation were superimposed on much faster circular flows, in which defects completed a circular orbit with a $\sim$10 s period (\fig\ref{figure3}D). In the vast majority of cases the boundary-induced defect nucleation preserved the existing handedness of the circular flows. However, on rare occasion reversal of flow handedness was observed (movie S1 at $\sim$4,300 seconds). Fast circular flows with $\sim$10 s periodicity are also seen in temporal evolution of the velocity fields obtained by using particle imaging velocimetry (PIV) on the fluorescently labeled microtubules (\fig\ref{figure3}C, movie S2). However, caution needs to be taken when interpreting these results since the PIV algorithm, which detects changes in light intensity, does not track dynamics that are tangent to uniformly aligned regions when the microtubules are densely labeled; thus such data is semi-quantitative.

To gain insight into the mechanisms underlying the doubly-periodic dynamics, we extracted active stresses from the evolving nematic director field. Hydrodynamic theory postulates that these stresses are proportional to the nematic order tensor $\mathbf{Q}$. Thus, the force that drives the autonomous flows is related to the gradients of the active stress, and regions of spatially varying nematic order are required to generate flows~\cite{Simha2002, Marchetti2013}. For extensile active nematics these forces point from high to low curvature. An experimentally measured map of the nematic director field yields an estimate of the active force, $\propto \nabla\cdot\mathbf{Q}$, and the component of this force that drives circulating flows, $\propto\hat{\mathbf{e}}_{\theta}\cdot\nabla\cdot\mathbf{Q}$ (\fig\ref{figure3}E). Our analysis assumes that activity is independent of microtubule density. The initial asymmetric spiral director field generated significant gradients of stress in the azimuthal direction, which in turn powered strong circular flows in the direction consistent with the hydrodynamic theory of extensile active nematics (\fig\ref{figure3}F, t=10 s). The evolution of the system towards a more tightly wound spiral reduced the magnitude of this force. The boundary induced defect nucleation correlated with the time point where the active force in the azimuthal direction reached a minimum. This suggests that the sufficiently strong flow alignment generated by the circular flows suppressed the defect nucleation that might otherwise form through the bend instability. The boundary induced defect nucleation could proceed only once the flows were sufficiently weak due to reconfiguration of the director field and the associated active force.

\begin{figure}
\includegraphics[width=\columnwidth]{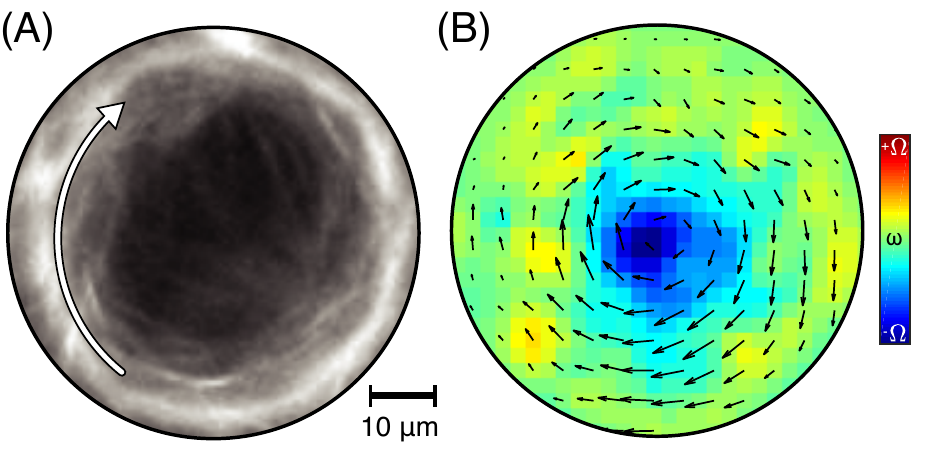}
\caption{\textbf{Active nematics in strong confinement} {\bf (A)} Circulating fluorescently labeled microtubules confined to a 60 $\mu$m diameter disk (Movie S4). {\bf (B)} Time-averaged (25 seconds) flow and vorticity fields (vorticity scale $\Omega=\pm$0.19 s$^{-1}$).}
\label{figure4}
\end{figure}

\subsection*{Transition to strong confinements} Doubly-periodic dynamics associated with the intermediate confinement regime persisted as the diameter was reduced to below 200 $\mu$m. However, for 100 $\mu$m confinements the dynamics exhibited new behaviors. In this regime the confinement size is comparable to the defect core size (Movie S3). Consequently, microtubule accumulation along the boundary became increasingly prominent. The center of the disk, if it contained microtubules at all, exhibited simple rigid body rotation driven by the boundary.  Reducing the confinement size further to 60 $\mu$m accentuated these behaviors (\fig\ref{figure4}, Movie S4). In this case the extending fibers no longer formed a continuous nematic field, but a fragmented ring at the boundary, while leaving large voids in the center. The temporal dynamics were no longer continuous and regular. Despite the loss of some structural order, we still observed fairly consistent circular rotations and intermittent buckling events. It is likely that the continuum approximation is no longer valid in this limit. Theory predicts that for small enough confinements, net circulation ceases because the active stresses are not strong enough to overcome the elastic distortions~\cite{Norton2018a}. We did not observed this prediction. 

\begin{figure*}
\includegraphics[width=\textwidth]{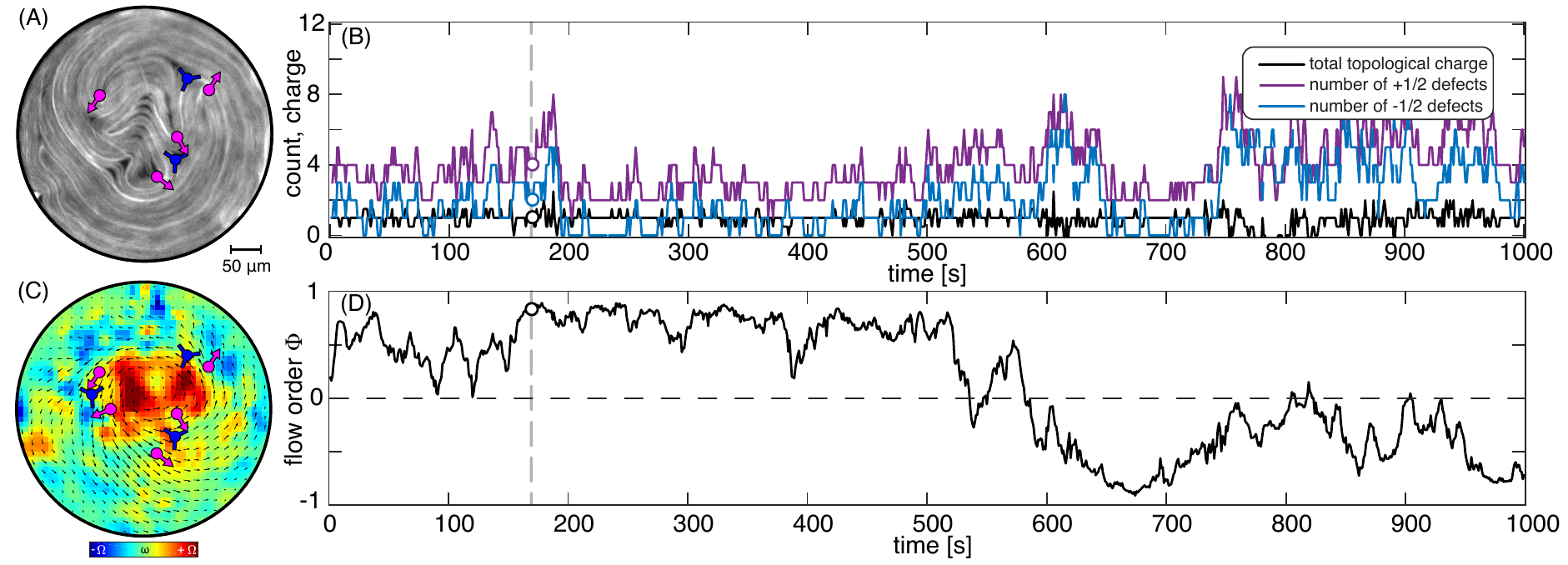}
\caption{\textbf{Active nematics in weak confinement} {\bf (A)} Active nematic in the regime where topological defects nucleate both in the interior and at the boundary; the \pdef defects are predominantly aligned along the flow direction.{\bf (B)} Time evolution of the number of \pdef defects (magenta), \mdef (blue), and the total topological charge (black). The total charge remains +1 except for short-time tracking errors. {\bf (C)} Spatial map of the instantaneous flow field and vorticity (vorticity scale is $\Omega$=$\pm$ 0.27 s$^{-1}$). {\bf (D)} Time evolution of the flow order parameter demonstrates that the circular flows with specific handedness persist for periods up to 10 minutes. The vertical line in (B) and (D) corresponds to the time in (A) and (C). Confinement diameter is 400 $\mu$m.}
\label{figure5}
\end{figure*}

\subsection*{Weak confinements} Active nematics confined to disks with diameter between 400 and 600 $\mu$m exhibited qualitatively new behavior compared to the intermediate and strong confinement regimes. In this regime defects nucleated both at the boundary and within the interior (\fig\ref{figure5}A). In addition to the two topologically required \pdef defects, at most times the system contained several additional defect pairs, whose number fluctuated anywhere between 2 and 10 (Fig.~\ref{figure5}B). Except for short term errors in defect tracking, the net charge of all defects added up to +1. Visual inspection revealed that these samples still exhibited coherent circular flows despite the presence of additional defects (Movie S5). To quantify this phenomena, we defined the signed order parameter that indicates the degree of flow circulation as $\Phi\left(t\right)=\left<\mathbf{u}\cdot\hat{\mathbf{e}}_{\theta}/\left|\mathbf{u}\right|\right>$, where $\mathbf{u}$ is the velocity field extracted from the PIV algorithm (Fig.\ref{figure5}C). $\Phi=\pm 1$ for system-scale coherent circular flows, while $\Phi=0$ indicates lack of net transport along the azimuthal direction~\cite{Theillard2017,Wu2017}. Temporal evolution of the flow order parameter confirms that weakly confined active nematics exhibited persistent circular flows that switched handedness on the time scale of tens of minutes (Fig.\ref{figure5}D).  

\begin{figure}
\includegraphics[width=\columnwidth]{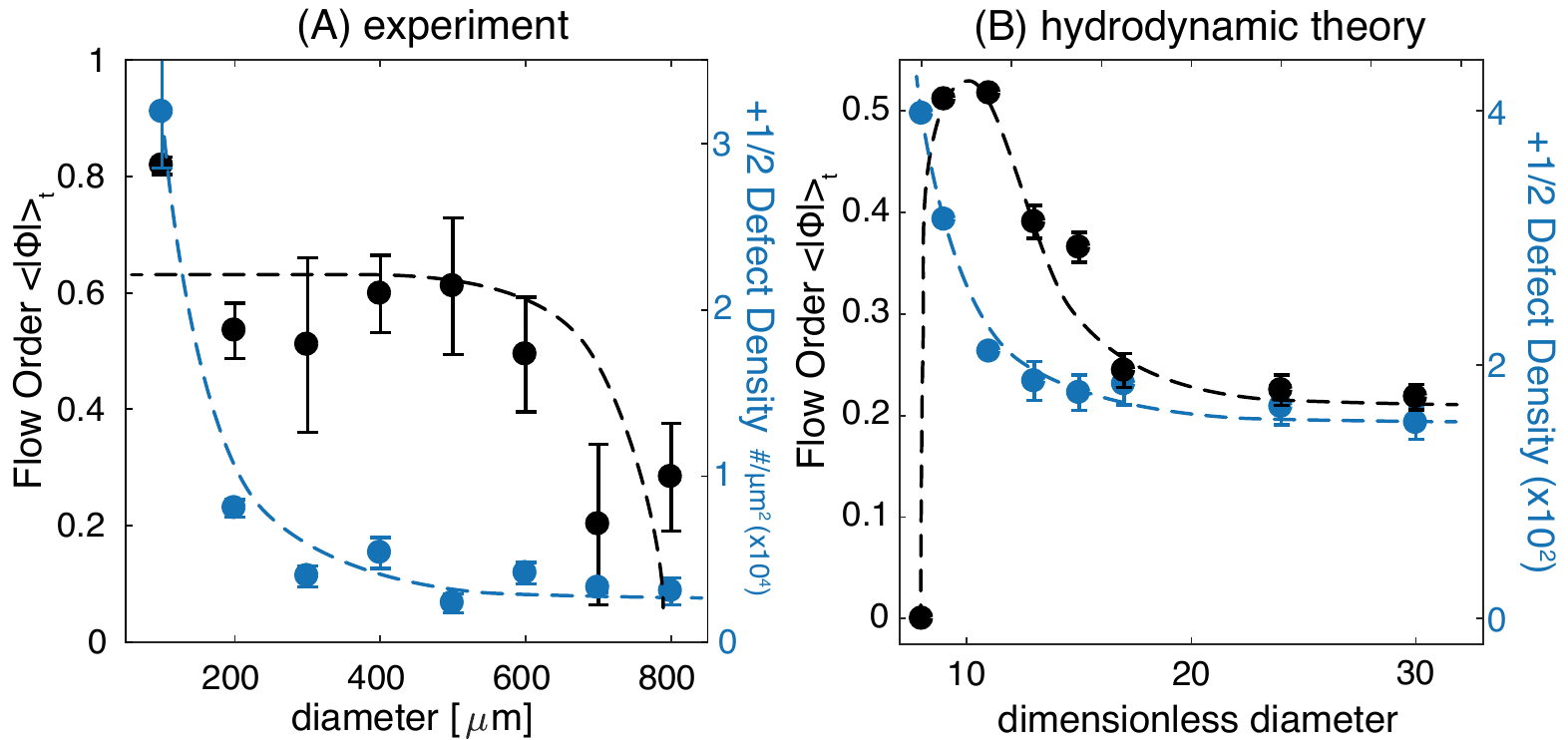}
\caption{\textbf{Dynamical and structural criteria for the onset of circular flows} {\bf (A)} Experimentally measured flow order parameter (black) and \pdef defect density (blue) is averaged over sample lifetimes and plotted as a function of the confining disk diameter. The onset of circular flows is observed for confinements below 600 $\mu$m, while the topological constraint does not modify the average defect density until the confinement is below 200 $\mu$m. Error bars represent standard error across different disks. {\bf (B)} Theoretical predictions of the flow order parameter and defect density plotted as a function of disk size for dimensionless activity $\alpha$=10. The onset of circular flows is tightly coupled to the increase of the defect density. Error bars represent standard error with $n=T/\tau$ measurements where $T$ is the simulation duration and $\tau$ is the average lifetime of a defect. Below a dimensionless diameter of $\sim$8 theory predicts a transition to a non-circulating ($\left<\left|\Phi\right|\right>_t=0$) state \cite{Norton2018a}.}
\label{figure6}
\end{figure}

\subsection*{Transition to bulk turbulence} So far we described the dynamics of active nematics in confinements ranging from 60 to 600 $\mu$m. For intermediate confinements, dynamics consisted of periodic boundary nucleation and the slow collapse of  the defect orbits superimposed on their rapid procession. Circular flows persisted for larger diameter disks in the weak confinement regime. Here defects nucleated both at the boundary and within the interior, and the nucleation dynamics were no longer deterministic but rather chaotic. The transition between these regimes was gradual. To compare the behaviors across all regimes we plotted the time-averaged flow order parameter $\left<\left|\Phi\right|\right>_t$ against the confinement size (\fig\ref{figure6}A). Self-organized circular flows had the same degree of flow order for confinements up to 600 $\mu$m. Beyond this size $\left<\left|\Phi\right|\right>_t$ dropped to zero, indicating the transition to bulk-like chaotic dynamics (Movie S6).  

\begin{figure}
\includegraphics[width=\columnwidth]{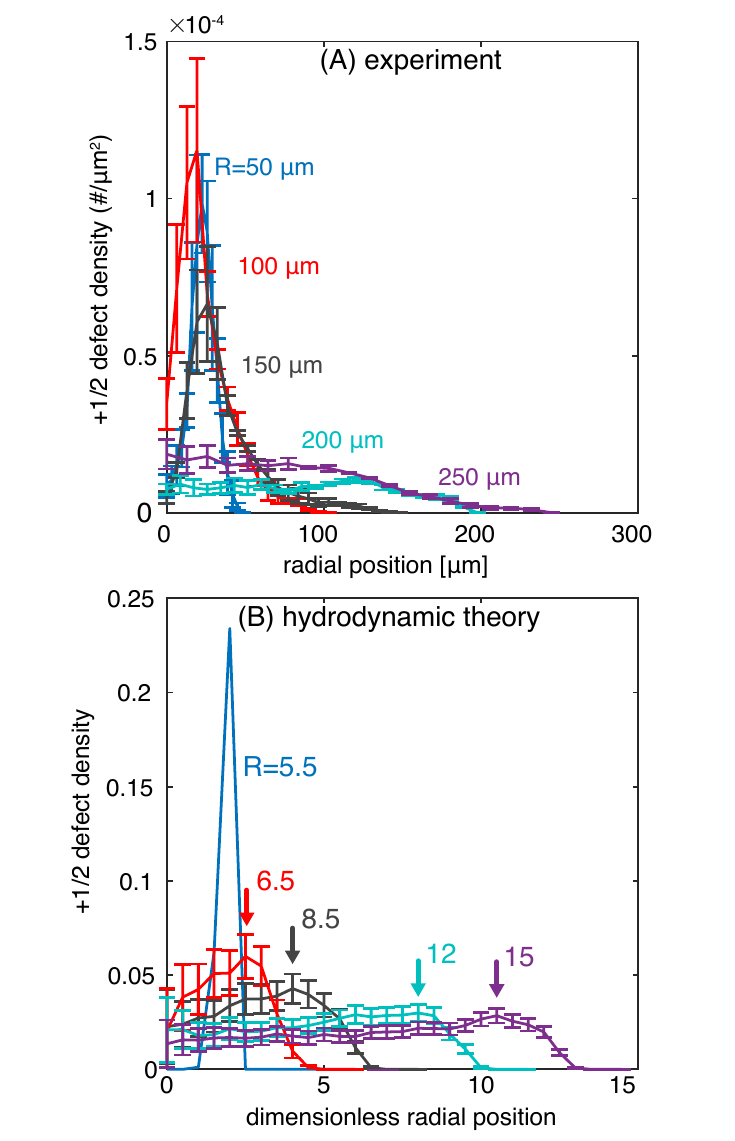}
\caption{\textbf{Radial distribution of defect densities across different confinement regimes} {\bf (A)} Time averaged radial distributions of \pdef defects plotted for confinements of different radii. For stronger confinements the topological constraint requires presence of two \pdef defects regardless of the energetic penalty, giving rise to a sharp peak in the defect density distribution at a well-defined radius. Error bars represent standard error across different disks. {\bf (B)} Theoretically calculated defect distribution plotted as a function of confinement size for dimensionless activity $\alpha$=10. In small diameter disks,  defects orbit at a preferred radius.  In larger diameter disks, defects tend towards uniform density in the disk interior with a local maxima (arrows) near the hard wall boundary \cite{Norton2018a}. Error bars represent standard error with $n=T/\tau$ measurements where $T$ is the simulation duration and $\tau$ is the average defects lifetime.}
\label{figure7}
\end{figure}

Motivated by the close relationship between topological defects and autonomous flows we examined how the defect density depends on the confinement size. In bulk active nematics, active stresses generate topological defects and their magnitude determines the preferred defect density. Reducing the confinement diameter leads to an increase of the effective density, since topological constraints ensure formation of at least two \pdef defects, even for confinements smaller than the average bulk defect spacing. Plotting the average defect density as a function of the confinement size reveals the expected increase for geometries less than $\sim 200$ $\mu$m (\fig\ref{figure6}A), which is comparable to the length scale extracted from the analysis of bulk active nematics~\cite{lemma2018statistical}. Therefore, from a structural perspective active nematics become unconfined above $\sim 200$ $\mu$m confinements. In contrast, the dynamical signature of confinement as manifested by the emergence of circular flows, persists for much weaker confinement, remaining even as the diameter is increased to $\sim 600$ $\mu$m (\fig\ref{figure6}A). We compared these findings to predictions of the hydrodynamic theory by determining how the defect density and the flow order parameter change with the confinement size~\cite{Norton2018a}. Theory predicts that the onset of the circular flows coincides with the increase in the effective defect density (Fig.\ref{figure6}B). Equivalently, circular flows appeared only when the system was confined below the defect spacing preferred by the bulk samples. Increasing the confinement size to the point where any additional defects appear immediately suppressed circular flows. This is in contrast to experiments where active nematics exhibited robust circular flows even in the presence of several additional defect pairs. Thus, theory predicts that the transition to bulk dynamics occurs by a one step transition characterized by concomitant changes in the defect density and the flow order, while in the experiment the structural and dynamic transitions occur for different confinements.

We examined an additional structural aspect of confined active nematics by plotting the time-averaged radial distribution of \pdef defect density for different confinement sizes (\fig\ref{figure7}A). For moderately confined active nematics, the topological constraints gave rise to a defect density distribution with a peak at a well-defined radial position. Intriguingly, the radial position of the peak did not significantly change, even as the confinement diameter increased from 100 to 300 $\mu$m. Beyond this diameter the peak in the radial density distribution disappeared. This drop coincides with the onset of the weak confinement dynamics in which defects proliferate (\fig\ref{figure5}). In this regime the measured probability distribution was flat within the interior of the disk, and decayed to almost zero upon approaching the boundary. The theoretical model captures both the sharp peak at strong confinements, and the flattening of the distribution as the confinement diameter is increased (\fig\ref{figure7}B, \cite{Norton2018a}). However, the details of the predicted defect distribution diverge subtly from the experiment; theory predicts that the maximum of the distribution moves outward with increasing confinement size, in contrast to experimental observations of a fairly large boundary exclusion zone with low defect density. This suggests that the dynamics that direct the \pdef defects towards the center at intermediate confinement are also relevant for larger diameters, transporting defects towards the interior of the disk.

\begin{figure}
\includegraphics[width=\columnwidth]{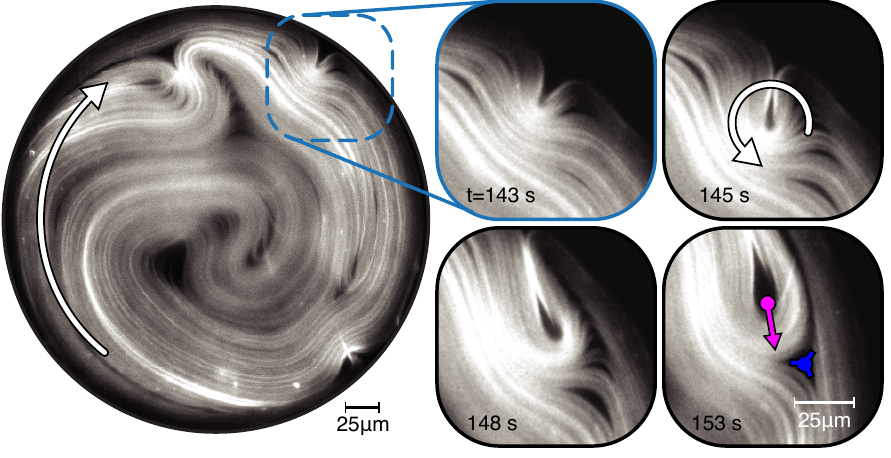}
\caption{\textbf{Dynamics of boundary-nucleated defects.} Structure of circularly confined defect-laden active nematic in a weak confinement regime (movie S7). Inset: A sequence of images showing a boundary induced defect nucleation and its subsequent rotation and alignment with the existing director field that reinforces the circular flow. Such dynamics produce long-lived compound \pmdef pairs featuring a \pdef defect that pushes its oppositely charged counterpart for a substantial time before annealing. Confining diameter, 300 $ \mu$m }
\label{figure8}
\end{figure}

Examining defect motion subsequent to the boundary nucleation provides some insight into the discrepancy between the theory and experiments. Initially, the instability grows perpendicular to the boundary. In isolation a nucleated \pdef defect would presumably continue to move towards the confinement center. However, existing circular flows and the associated nematic director field rapidly reorient the newly created defects in the azimuthal direction (\fig\ref{figure8}). Defects with such alignment generate additional active stresses that further stabilize circular flows. Furthermore, evolving dynamics lead to the creation of a compound defect structure composed of a \pdef defect with a \mdef defect that have substantial lifetime before annealing (t=153 s~\fig\ref{figure8}, Movie S7). Together, the defect pair acts as an active wedge that further stabilize the circular flows. 

\subsection*{Non-circular geometries} 
\begin{figure*}
\includegraphics[width=\textwidth]{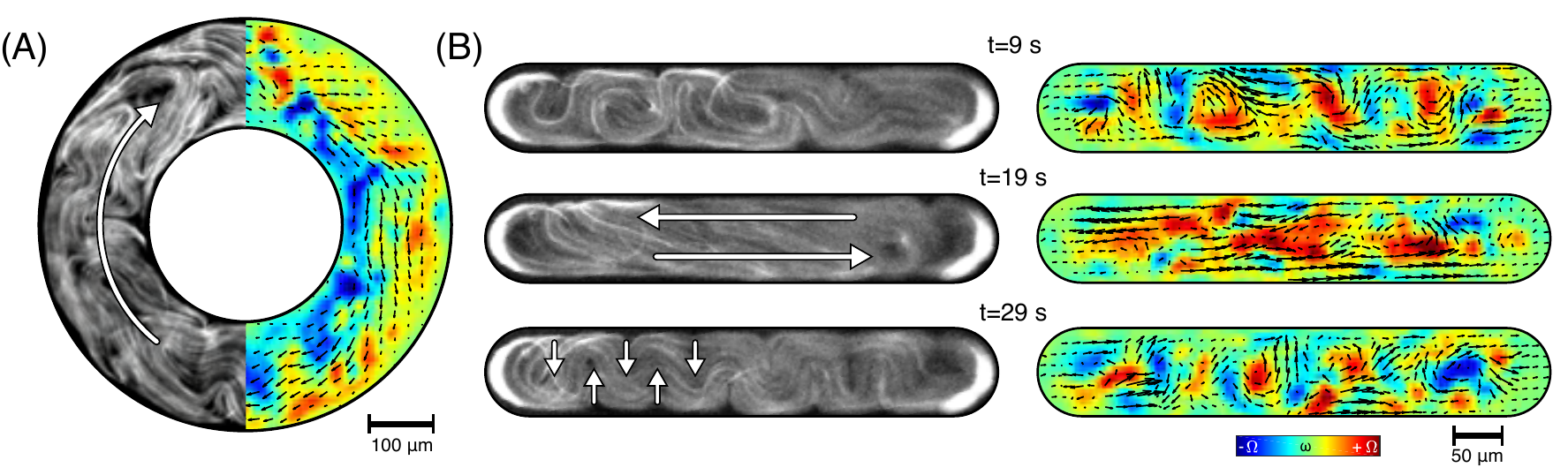}
\caption{\textbf{Active nematics in complex geometries (A)} Active nematic confined in a 150 $\mu$m wide annulus. Fluorescently labeled microtubules are shown on the left and the flow and vorticity field extracted from the PIV analysis is shown on the right (vorticity scale is $\Omega$=$\pm$ 0.3 s$^{-1}$). Temporal evolution is shown in movie S9. {\bf (B)} Active nematic dynamics in rectangular confinements. Left column shows evolving structure of the nematics field. Right column shows the velocity field and the destruction and re-creation of a single row of vortices with alternating sign (vorticity scale is $\Omega$=$\pm$ 0.4 s$^{-1}$). Temporal evolution is shown in movie S10.}
\label{figure9}
\end{figure*}

So far we described dynamics of circularly confined active nematics. However, the developed experimental methods are more general and can be used to study dynamics in diverse confined geometries. To demonstrate their versatility we confined active nematics within a disk with an inward notch, an annulus, and rectangular boxes of different lengths and widths. A chiral notch at the confining boundary of a disk controls the spatial location of defect nucleation (Movie S8). It also ensures that autonomous flows have only one handedness, which is reminiscent of the methods that control the flow of confined 3D isotropic active fluids~\cite{Wu2017}. Our experiments showed that confining active nematics within an annulus provides an alternate method of engineering coherent long-ranged flows (\fig\ref{figure9}A, Movie S9). It remains an important question to determine how the annular width and curvature determines the structure of the self-organized flows. We also observed intriguing dynamics in active nematics confined within a rectangular box with a large aspect ratio (\fig\ref{figure9}B, Movie 10). For the majority of the time the dynamics of this system are chaotic. However, occasionally we observed an organized state in which defects collectively and simultaneously nucleated at the boundary and subsequently propagated into the interior.  

\section*{Discussion and Conclusion} 
Persistent circular motion has been observed in a wide range of confined active matter systems ranging from clusters of motile eukaryotic cells, and dense bacterial suspensions to synthetic motile colloids~\cite{doxzen2013guidance,Wioland2013,bricard2013emergence,bricard2015emergent, tee2015cellular,Wu2017}. All of these systems are characterized by a single time scale related to the frequency of the rotational dynamics. Our observations confirm the universality of circular flows in the confined active matter. However, when compared to previous experiments and to theoretical predictions, we observed two ubiquitous yet unexplained phenomena. First, in the intermediate confinement regime the nematic director field formed a double spiral that evolved over time through the multiple procession cycles of \pdef defects. The slow collapse of the \pdef defect orbits drove the system towards a tightly wound, nearly circularly symmetric configuration. Once sufficiently circular, the material was reconfigured by a boundary-induced defect nucleation, initiating a subsequent cycle of the doubly-periodic dynamics. Second, as a function of the increasing diameter we observed a two-step transition from the doubly-periodic dynamics to bulk turbulence. Increasing confinement beyond a critical diameter led to a profusion of additional defects that nucleated both at the boundary and within the interior. The density of defects approached the value measured in bulk materials, yet flows remained ordered. Circular flows ceased and bulk chaotic dynamics emerged only after increasing confinement diameter further to the second larger value. 

Existing continuum theories of active nematics predict suppression of spontaneous flows below a critical length scale and activity threshold, density flux from high to low director curvature, chaotic dynamics at high activity and large length scales; and regular circulating flows in the intermediate regime~\cite{Woodhouse2012, Putzig2015, Gao2015d, Gao2017, Shendruk2017, Norton2018a, chen2018dynamics, theillard2017geometric}. While these models are not identical, nor are the applied boundary conditions, they all fail to predict both the doubly-periodic dynamics at intermediate confinement and the two-step route to bulk-like turbulence. Agent-based models of active nematics might overcome limitations of continuum models by explicitly modeling particle-particle interactions; these models predict buckling instabilities that produce motile defects \cite{Decamp2015,Joshi2018,Li2018,Yaman2018} but have not considered the impact of boundary conditions essential for reproducing experimental observations. 



Dynamics similar to those described here have been observed in active nematics assembled on smectic focal-conic domains which provide soft circular confinement~\cite{guillamat2017taming}. In both cases there is pronounced migration of microtubules from the high-curvature interior to the lower-curvature boundaries. However, the dynamics that disrupt the circular structure in the two systems are distinct. The smectic-confined circular nematics lose stability along the entire periphery with a well-defined wavelength. In contrast, the hard-wall boundary prevents outward buckling, causing nucleation to occur as a distinct, spatially localized event that produces a single \pmdef defect pair.

Experiments suggest two possible mechanisms that control the doubly periodic dynamics. One possibility is that the build-up of the microtubule density at the boundary increases local active stresses. Once these reach a critical magnitude the generic nematic instability develops, leading to boundary induced defect formation. Another possibility is that the defect nucleation is suppressed by the persistent circulating flows that tend to align the director azimuthally. As the defects wind the double spiral tighter, there is a marked decrease in the total active force generated in the azimuthal direction, which leads to the slowdown of the associated flows. Once they decrease below a certain threshold, the flows no longer suppress defect nucleation. This disinhibition initiates another cycle of defect dynamics. These two mechanisms are not mutually exclusive; both increased activity and weakened suppression of defect nucleation could be experimentally relevant to controlling the temporal onset of boundary induced defect nucleation events. 

\begin{figure}
\includegraphics[width=\columnwidth]{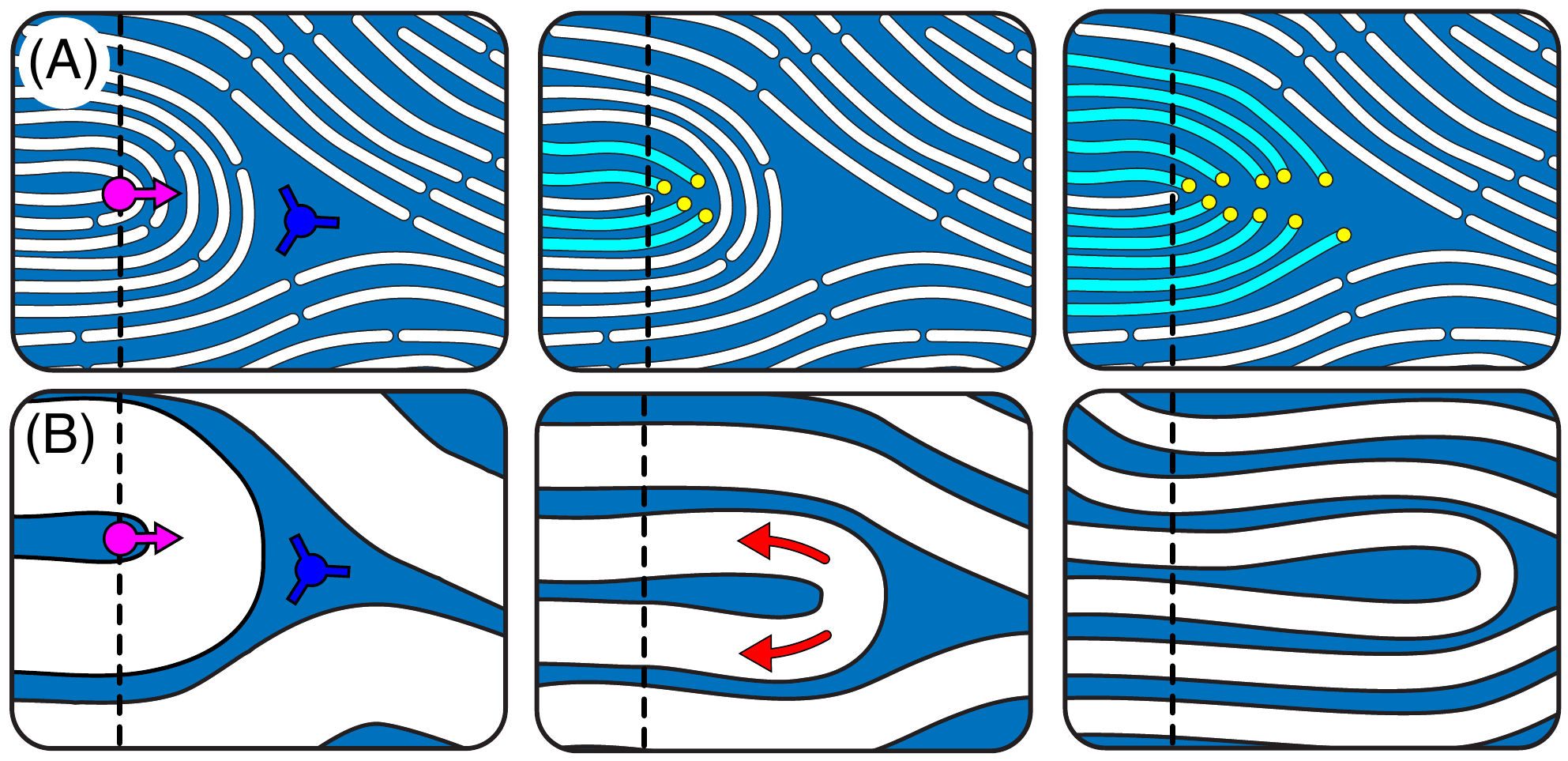}
\caption{\textbf{Mechanisms of defect propulsion and annihilation:} In the microtubule active nematic defect propulsion can occur through a combination of thinning and fracture mechanisms. \textbf{(A)} Constituent fibers reorient and fracture allowing the \pdef defect core to move forwards. The fracture driven motility does not require material transport. \textbf{(B)} In the regime where constituents cannot easily reorient, ordered fibers lying in front of the \pdef defect prevent its forward movement. Extensile active transport within the nematic along director lines (red arrows) causes elongation and thinning that both propels the defect pair forward and brings them closer together. Their annihilation rate is limited by the rate at which the extensile flows can displace the materials in front of the \pdef defect. Dashed line indicates the initial defect position.}
\label{figureLCpolymer}
\end{figure}

Our observations of defect dynamics further suggest that microtubule based active nematics are composed of continuous material lines that can be distorted but not crossed by defects. This is evidenced by defect reorientation dynamics (\fig\ref{figure2}A and \fig\ref{figure8}). Driven by molecular activity, the nematic undergoes volume-conserving extension along the director field and contraction in the direction perpendicular to the nematic field (cyan line in \fig\ref{figure2}A). In general, an active \pdef defect can move forward through two mechanisms: one that involves no net material transport but local reorientation of the nematic director field (\fig\ref{figureLCpolymer}a), and one that requires active expulsion of material in front of the defect along the director field (\fig\ref{figureLCpolymer}b). The former, while realizable in thermotropic liquid crystals and perhaps even bacterial active nematics \cite{Li2018}, is supressed by the long constituents of the microtubule active nematic. Given the constraints imposed by the director-defined material lines, we observe a preference for the latter mechanism. Similar considerations are also important for the mechanisms of defect annihilation. In low molecular weight nematics simple reorientation of the nematic director can lead to defect annihilation. In contrast, in microtubule active nematics the filaments between the annihilating defects have to be expelled through active transport of constituent fibers driven by extensional flows. This might explain the long-lived nature of compound defect pairs (t=153 s~\fig\ref{figure8}). Current continuum liquid crystal models lack the constraints on material motion needed to predict the observed defect interactions~\cite{Giomi2013, tang2018theory, shankar2018defect, Cortese2018}.
 
In summary, we demonstrated that confinement effectively transforms the chaotic dynamics of bulk 2D active nematics in multiple ways, inducing the formation of self-organized doubly-periodic spatiotemporal patterns. These observations pave the way to controlling dynamics of active nematics by tuning the confinement geometry. However, predictive control of the emergent patterns requires both further experimental explorations, as well as development of theoretical models that more accurately account for the experimental realities of  defect nucleation and interactions. More broadly, the dynamics of our reconstituted minimal system are evocative of the phenomena observed in living organisms, which frequently exhibit complex spatiotemporal patterns that emerge on lengthscales and timescales that are significantly larger than the size and the lifetime of the biochemical constituents~\cite{Brugues:2014aa}. Using kinesin motors that take an 8 nanometer sized step every ten milliseconds, we demonstrated self-organized spatiotemporal patterns with lifetimes of minutes on lengthscales of hundreds of microns. Understanding the fundamental organizing principles that drive the dynamics of active nematics might reveal general strategies for rationally engineering adaptable and dynamical synthetic materials with biomimetic capabilities. 

\section*{Methods}

\subsection*{Microtubule based active nematics}
We used active nematics comprised of three components: GMPCPP stabilized microtubules with an exponential length distribution with a mean of 1-2 $\mu$m \cite{Decamp2015, guillamat2016control} labeled with Alexa 647 dye, biotin labeled kinesin-motors bound into multi-motor clusters by tetrameric streptavidin~\cite{nedelec1997self}, and a depletion agent poly(ethylene glycol), that induces microtubule bundling while still allowing for their relative sliding~\cite{needleman2004synchrotron, hilitski2015measuring,ward2015solid}. The composition and preparation of the active mixture is the same for all experiments. Kinesin clusters simultaneously bind multiple microtubules in a bundle, and depending on their relative polarity drives the bundle extension \cite{Sanchez2012}. Purified microtubules were flash frozen, stored at -80 $^{\circ}$C, and rapidly thawed before use; they were kept at room temperature (20-25 $^{\circ}$C) throughout the experiment. Following previous work, we sediment the extensile bundles onto a surfactant stabilized oil-water interface where they assemble into a dense quasi-2D active nematic film. The aqueous suspension also contained an ATP regenerating system to increase the experimental life-time and an anti-oxidant system to prevent photobleaching~\cite{Sanchez2012,Decamp2015}. Depending on the source of the kinesin motor proteins and microtubules the dynamics of active nematic can vary significantly. Similarly we found that these parameters greatly affected the frequency of circulating active nematics. However, they did not  affect the confinement size where dynamics transitions between strong, intermediate and weak regimes. 

\subsection*{Fabricating micro-chambers for confinement} An array of circular micro-chambers with diameters that ranged from 60 - 800 $\mu$m were fabricated on glass microscope slides using standard photolithography techniques, (\fig\ref{figure1}A). The $3\times4$ inch microscope glass slides were pre-cut to $3\times3$ inch using a diamond tip scriber, in order to fit them in the vacuum chuck of the mask aligner. The slides were then rinsed and sequentially sonicated with (1) hot water containing 0.5\% detergent (Hellmanex III, Hellma Analytics), (2) then with ethanol, and (3) finally with 0.1 M NaOH. In order to enhance SU8 bonding to glass, the slides were immersed in a silane mixture consisting of 98\% deionized water, 1\% acetic acid and 1\% (3-aminopropyl) trimethoxysilane (Sigma Aldrich) for 20 minutes. Slides were then rinsed with deionized water, dried with N$_{2}$ gas and placed on a hot plate at 180 C for 15 minutes to remove moisture. Once  cooled, photoresist SU8 3025 (MicroChem Corp.) was spin-coated on the slides at 1400 rpm  (Headway Research). Subsequent standard photolithography processing steps, including soft baking for 20 mins at 95 C, UV exposure through a photomask with 180 mJ/cm$^{2}$ and post exposure baking for 10 minutes at 95 C. Photoresist development was carried out on the slides followed by hard-baking for 30 minutes at 180 C. The depth of the micro-chambers was $\sim30 \mu$m as measured by an optical profilometer.   

\subsection*{Assembling the flow cell and confining the active nematics} In order to prevent non-specific adsorption of microtubules and motors and to create a hydrophilic surface, the lithographed microscope glass slides and no$.$ 1.5 coverslips were plasma etched and coated with polyacrylamide as described elsewhere~\cite{Lau2009}. Subsequently, each slide was rinsed and dried with N$_{2}$ gas immediately before use. Polyacrylamide treated lithographed slides and cover slips were assembled into a flow cell using $\sim 100 \mu$m  thick double-sided transfer tape (3M 93005LE)  with laser-cut flow channels. HFE7500 oil (3M) with 1.8\% (v/v) Fluoro-surfactant (RAN Biotechnologies) was flowed into the cell. Subsequently, the aqueous active mixture was flowed through the channel, displacing the oil-surfactant mixture from all but the hydrophobic SU8 micro-chambers. The flow channel was then sealed with optical UV curable glue and centrifuged at 1000 rpm for 10 min using a Sorval Legend RT (rotor 6434) to drive the bundled microtubules onto the surfactant-stabilized oil-water interface. Within each sample on average about 20\% of disks had well-defined circularly confined active nematic. Fluorinated oil often wetted areas of the SU8 photoresist, leading to the regions that formed bulk-like unconfined active nematics. However, since there were several hundred micro-chambers per slide and less than 10 could be imaged during the lifetime of the active nematic, the 20\% success rate was sufficient.

\subsection*{Imaging confined 2D active nematic} The confined active nematic was imaged using a wide-field fluorescent Nikon Eclipse Ti-E microscope with an Andor Clara camera controlled by Micromanager open-source software. The dynamics of quasi 2D active nematics is dependent on the thickness and concentration of microtubule film, the product of which is proportional to the optical retardance, which was determined with an LC-PolScope module~\cite{Decamp2015}. Birefringence of the SU8 photoresist limited our ability to quantify the retardance of the confined active nematic. In order to avoid spillover of signal from the SU8 periphery, the field aperture was partially closed to image about three quarters of the confined area. These measurements determined the upper bound of the retardance of the nematic to be about a nanometer, which is close to the retardance resolution limit.  Because of the small ratio of signal to noise, we could not  quantitatively compare the retardance of the contents of the disks across experiments.

\subsection*{Image Processing of the Director Field} In contrast to previous studies, the high birefringence of SU8 photoresists precludes the use of the LC-Polscope  to directly quantify the director field~\cite{Decamp2015}. Instead, the director field was extracted from fluorescence images. The nematic order tensor, $\mathbf{Q}=\mathbf{n}\otimes\mathbf{n}-\frac{1}{2}\mathbf{I}$, was constructed by finding the direction $\mathbf{n}$, perpendicular to the principal direction of the structure tensor of the image $\nabla I \otimes \nabla I$, where $I$ is image intensity pixel value~\cite{ellis2018curvature}. The degree of order, $s=2\Tr{\mathbf{\left<Q\right>}^2}$, was found by averaging $\mathbf{Q}$ over a small region. To avoid assigning nematic order to regions that are largely devoid of microtubules, the displayed scalar order parameter field $S$ was weighted by the image intensity ($I$) if the pixel value was below 10\% of the maximum ($I_{\mbox{max}}$); $S = s*I/I_{\mbox{max}}$, with $s$  the unweighted order parameter. Defects in the director field were found using a previously developed algorithm~\cite{Decamp2015}. Defect tracking errors for the data set represented in \fig\ref{figure5} are shown in the electronic supplement, \fig S2.

\subsection*{Processing of the Flow Field} The velocity field $\mathbf{u}$ was found by applying particle image velocimetry analysis to the fluorescently labeled microtubules using the MatLab plug-in PIVlab v1.43 \cite{Thielicke2014pivlab}. In typical PIV applications, the flow is quantified by following sparsely labeled, small, isotropic tracer particles. Since these conditions would decrease the quality of the measured director field, PIV analysis is performed on densely labelled microtubules. Consequently, velocity components parallel to the director are under-sampled.  For example, since the nematic is nearly parallel along the boundaries, we expect the PIV analysis to underestimate the tangential flows in these regions and consequently, the order of tangential flow ordering $\Phi=\left<\mathbf{u}\cdot\hat{\mathbf{e}}_{\theta}/\left|\mathbf{u}\right|\right>$ discussed in the text. We were still able to detect circulation in many cases because \pdef defects dominate motion. They feature a nematic field perpendicular to the direction of their motion, which is well captured by the PIV algorithm.

\section*{Acknowledgements} We acknowledge useful conversations with Michael Hagan, Aparna Baskaran, and Stephen DeCamp. Experimental work on circular confinements was supported by Department of Energy of Basic Energy Sciences, through award DE-SC0010432TDD. Development of the microfluidic confinement technology and extension to non-circular geometries was supported by Brandeis MRSEC through grant NSF-MRSEC-1420382. We also acknowledge the use of the MRSEC optical, microfluidic and biosynthesis facility supported by NSF-MRSEC-1420382. Development of the computational model of confined active nematics was supported by NSF-DMR-1810077.

%

\end{document}